\documentclass[a4paper,11pt]{article}
\pdfoutput=1 

\usepackage{jcappub} 

\usepackage[T1]{fontenc}
\usepackage[latin9]{inputenc}
\usepackage{float}
\usepackage{bm}
\usepackage{dcolumn}
\usepackage{color}
\usepackage{amsmath}
\usepackage{amssymb}
\usepackage{graphicx} 
\usepackage{hyperref}
\hypersetup{colorlinks,citecolor=blue}
\usepackage[normalem]{ulem}

\begin{document}

\title{Does History Repeat Itself? \\Periodic Time Cosmology} 

\author[a,b,c]{Elizabeth Gould} 
\author[b,c,d]{Niayesh Afshordi}

\affiliation[a]{The McDonald Institute and Department of Physics, Engineering Physics, and Astronomy, Queen's University, Kingston, Ontario, K7L 2S8, Canada}
\affiliation[b]{Department of Physics and Astronomy, University of Waterloo, Waterloo, ON, N2L 3G1, Canada}
\affiliation[c]{Perimeter Institute for Theoretical Physics, 31 Caroline St. N., Waterloo, ON, N2L 2Y5, Canada}
\affiliation[d]{Waterloo Centre for Astrophysics, University of Waterloo, Waterloo, ON, N2L 3G1, Canada}
 
\emailAdd{egould@uwaterloo.ca}
\emailAdd{nafshordi@pitp.ca} 
 
\date{\today}

\begin{abstract}
{It has been suggested that the cosmic history might repeat in cycles, with an infinite series of 
similar aeons in the past and the future.
Here, we instead propose that the cosmic history repeats itself exactly, constructing a universe on
a periodic temporal history, which we call {\it Periodic Time Cosmology}.  In particular, the primordial power spectrum, convolved with the transfer function throughout the cosmic history, would form the next aeon's primordial power spectrum. 
By matching the big bang to the infinite future
using a conformal rescaling ({\it a la} Penrose), we uniquely determine the primordial power spectrum, in terms of the transfer function up to two free parameters.
While nearly scale invariant with a red tilt on large scales, using Planck and Baryonic Acoustic Oscillation observations, we find the minimal model is disfavoured compared to a power-law power spectrum at $5.1\sigma$. 
However, extensions of $\Lambda$CDM    cosmic history change the large scale transfer function and can provide better
relative fits to the data. For example, the best fit seven parameter model for our Periodic Time Cosmology, with $w=-1.024$ for dark energy equation of state, is only disfavoured relative to a power-law power spectrum (with the same number of parameters) at $1.8\sigma$ level. Therefore, consistency between cosmic history and initial conditions provides a viable description of cosmological observations in the context of Periodic Time Cosmology.}
\end{abstract}

\maketitle

\section{Introduction}


The preponderance of cosmological observations, ranging from big bang nucleosynthesis and cosmic microwave background to large scale structure and supernovae Ia can be explained by a nearly 
isotropic, homogeneous, and flat cosmological spacetime with small inhomogeneities that start nearly scale 
invariant. However, there is not enough time for all regions of the 
universe we see to have come in causal contact with each other to explain the observed homogeneity and correlations, otherwise known as the horizon problem.
What is the origin of the observed homogeneity of the universe? 
What about the small near-scale invariant fluctuations on top of this homogenous background?

Our current empirical model for the universe, known as $\Lambda$CDM    cosmology, matches (almost) all the observational data very well with six free parameters \cite{Bennett:2012zja, planck2018cosmo}.
Out of these six parameters, four describe the cosmic content and history, while two quantify the spectrum of inhomogeneities as a power-law, shortly after the big bang. This primordial power spectrum, $\mathcal{P}(k)$ is approximated by 
\begin{equation}
\mathcal{P} \left( k \right)=A_{0}\left(\frac{k}{k_{*}}\right)^{n_{s}-1}, \label{power_law}
\end{equation}
where $A_0$ and $n_s$ are unknown parameters, and $k_*$ is an arbitrary ``pivot scale''.
If we consider this as a Taylor expansion of $\log {\cal P}$ in terms of $\log k$, adding more terms to the expansion does not give a sufficiently better fit to 
warrant additional parameters \cite{planck2018cosmo}.

The most popular explanation of this structure for primordial power spectrum is the slow-roll inflationary paradigm \cite{Linde:1981mu, Mukhanov:1990me}.
Inflation posits that in the very early universe, the cosmic expansion went through a 
phase of rapid acceleration, which tends to homogeneize and flatten our observable universe \cite{Guth:1980zm}, and can produce near-scale invariant metric fluctuations \cite{Mukhanov:1990me}. While inflationary paradigm has been very successful due to its simplicity and its ability to fit cosmological observations, it has been criticized for various reasons \cite{Brandenberger,Turok,infpr1,infpr2}, not the least of which is that it might be impossible to falsify \cite{InflatSch}.  

Due to these problems, alternative suggestions for the early universe have been constructed by taking 
seriously the question: ``What came before the big bang?''.
If the big bang was not the beginning of time (or if there was no big bang), then there 
must have been a previous phase of the universe. This possibility typically involves 
replacing the big bang singularity with a bounce, where before the big bang, the universe 
was larger, and presumably contracted before expanding again (e.g., \cite{Gasperini:1992em}). This could allow for the 
necessary time for the universe to become correlated, eliminating the need for inflation. 

In addition, we have now discovered a non-vanishing cosmological constant, 
or dark energy. A late-time dark energy dominanated period before the bounce could cause 
similar effects to inflation \cite{ekpyro}. This implies one can construct a model where the conditions 
of the end of our universe match the expected conditions of the beginning. If the previous phase 
matches the end of the current universe, the model thus formed would
be cyclic, where the universe goes through a series of similar histories. Models based on 
this insight include the ekpyrotic cyclic model \cite{ekpyro, ekpyroO} and Penrose's conformal cyclic 
cosmology (CCC) \cite{ccc}.

While all these models imagine the universe to repeat itself, potentially infinitely 
many times, the geometric structure of general relativity (GR) allows for another possibility:
Since the spacetime manifold need not be flat globally, it is possible to construct a 
manifold for which, even without faster than light signalling, some events may be able
to influence their own causal past \cite{godel}. 
These spacetimes are said to have closed timelike curves (CTC). 

In such a spacetime manifold, it would not be possible to define generic initial conditions 
and simply evolve the system from the past to the future (i.e. solve a Cauchy problem in the standard way)  as is typically done in physics. Because of this, 
regularly manifested by the ``grandfather paradox'', considering a universe with CTCs is 
often avoided. For instance, causal set theory \cite{causalsets} and shape dynamics \cite{shapedynamics} both state removing these as part of their strengths.
In addition, with the standard energy conditions, added to require ``reasonableness'' 
and prevent runaway particle creation, CTCs cannot occur in general relativity \cite{grcausal}. 

However, it has been shown that when there is a CTC in a system, there exist self-consistent particle solutions 
for any choice of initial conditions before the CTC \cite{grgconf,ctc1,nov}. Moreover, solving dynamical equations based on an evolution from the past to the future is 
not the only way to do physics. The principle of least action, for instance, provides a description 
of what happens between two points by fixing the beginning and the end, from which Novikov's self-consistency condition can be derived \cite{Lctc1,Lctc2}.

So, if there is a repeating cosmic history, is it possible that the reason for this repetition 
is that the future is the past, and that we live on CTCs instead of time being infinite?
Most cyclic models avoid this possibility for various reasons, one being that this 
constraint is difficult to work with. However, if a model is more constrained, it would  
potentially be more predictive. And if the model succeeds with the additional constraint, 
this would be more significant.

With this suggestion, we construct a cosmological model based on the constraint of {\it temporal periodicity}. 
We ask if we can take advantage of this constraint to construct a testable form 
of the primordial power spectrum.

It turns out we can. To do 
this, we take advantage of the concept of conformal matching of CCC. 
Penrose suggested that in the infinite future, most to all matter would 
evaporate to radiation. Without massive particles, time no longer has any meaning and the 
structure is conformally 
invariant, allowing an arbitrary rescaling. The big bang period was radiation dominated, and thus 
was also effectively scale invariant early on. Therefore, given the conformal invariance of our effective theories near big bang, and infinite future, these can be matched to each other, as the singularity only appears on the conformal factor
\cite{ccc}.

As the superhorizon metric perturbations freeze at both big bang, and infinite future, they could be matched across the transition.
If we match larger comoving scales from the previous cycle to smaller comoving scales in the next 
cycle, the metric power spectrum should transfer from one cycle to the next. Therefore, we do not need to construct a 
mechanism to produce metric fluctuations. Instead, as we shall see, the effects of matter transfer function from 
the previous cycle would produce a slight red tilt on an otherwise scale invariant power spectrum on large scales. 

In this paper, we examine if this power spectrum produces a satisfactory fit to 
observations such that 
a model which recycles its power spectrum at each cycle is viable. 
In Section \ref{CTC_model}, we develop the periodic time cosmology (PTC) model described here. In Section
\ref{data}, we examine how well this model can fit the data, compared to the standard power-law power spectrum. In Section \ref{discuss}, we discuss 
the physical implications of our results, and Section \ref{conclude} concludes the paper.

\section{Model Construction}\label{CTC_model}

\subsection{Finding the Primordial Power Spectrum}\label{ss:ppsd}

We desire to construct a periodic model for which the condition of temporal periodicity
will constrain the system, so that we can make predictions based on this construction.
What can we observe from the previous cycle? In the cosmological data, we see the 
primordial power spectrum as what comes through from the earliest times, as an input to 
our model. It must have either come from the previous cycle or have been formed at 
early times. We will take it as coming from the previous cycle. This means that in 
the infinite future, there is a power spectrum of metric perturbations at some scale that transfers over into our past. 

If the evolution of the universe {\it were not to affect the power spectrum}, there are two 
forms of solutions for which the matching of the infinite future to the early universe 
would preserve. First, if comoving scales are matched to themselves, any power spectrum 
would work. This produces no restriction on the form of the power spectrum, but would 
require any evolution of the power spectrum to either be periodic or to reverse itself. 
Since we do not expect this and are looking to restrict the form of the power spectrum, 
this case is uninteresting for our analysis.

Let us now define a general conformal rescaling across the infinite future/big bang transition, for the spatial metric $g_{ij}$ of uniform density hypersurfaces: 
\begin{equation}
g_{ij}(\alpha {\bf x})|_{\rm big~bang} \propto g_{ij}({\bf x})|_{\rm infinite~future},\label{matching_naive}
\end{equation}
where $\alpha$ is a constant factor. In this case, the power spectrum is matched across the transition with a fixed, non-unitary shift 
in the comoving scales. Any scale invariant primordial 
power spectrum would be preserved by any such matching. More generally though, a power spectrum
periodic in $\ln\left(k\right)$ with a period of $\ln(\alpha)$ will match this condition.

However, we know the power spectrum {\it does} evolve from its primordial shape as the 
universe evolves. For example, the CMB anisotropies we see now is determined by the power spectrum after it has 
lost power due to acoustic oscillations and Silk damping on small scales \cite{silk,SZbao}. This loss in power is defined by the 
transfer function $T\left( k,t\right)$, which can be calculated 
numerically \cite{Lewis:1999bs,Howlett:2012mh,camb_notes,Seljak:1996is,Zaldarriaga:1997va}, or approximated by a fitting function \cite{eisenhu} in $\Lambda$CDM    cosmology. It is defined 
such that it is normalized to 
\begin{equation}\label{eq:normT}
\lim_{k\rightarrow0}T(k, t )\rightarrow1.
\end{equation} 
To fit current observations, we only need to focus on the scalar adiabatic perturbations, quantified by the Bardeen variable $\zeta$
 \cite{Mukhanov:1990me}, which denotes the conformal fluctuations of the spatial metric on superhorizon scales in the uniform density gauge: 
\begin{equation}
g_{ij}({\bf x}) \propto [1+2\zeta({\bf x})] \delta_{ij},
\end{equation} 
and is related to the conformal Newtonian metric perturbations by
\begin{equation}
\zeta = \psi - \frac{H}{\dot{H}} \left(H\phi+\dot{\psi}\right).
\end{equation}

\begin{figure}
\includegraphics[width=\textwidth]{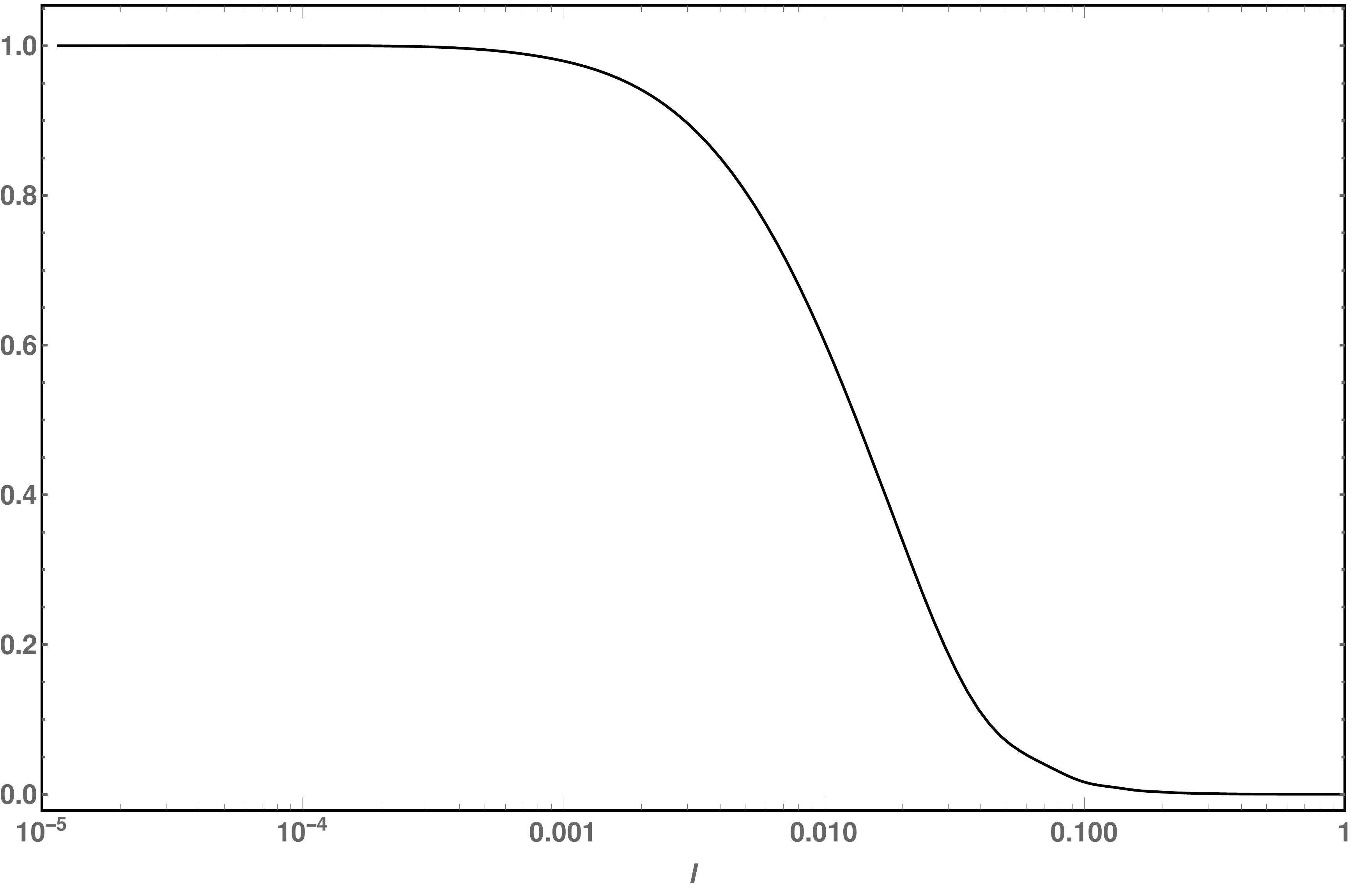}

\caption{\label{fig:trans1}
A plot of a sample transfer function evaluated at present time. The transfer function is normalized to one at the largest scales. This was calculated numerically in CAMB \cite{Lewis:1999bs,Howlett:2012mh} using values of $\Omega_b$, $\Omega_c$, $\tau$ and $100\theta$ from the best fit values for the standard $\Lambda$CDM fit to Planck 2018, using TT+TE+EE+lensing+BAO \cite{planck2018intro,planck2018cosmo,Ade:2015xua,Planck2016i,Ade:2013nlj,Ade:2015zua,Aghanim:2015wva,Ade:2015fva,Ade:2013zuv,Planck:2013kta,Ade:2013mta,Bennett:2012zja}. The transfer function we chose was the scale invariant $\zeta$ \cite{Mukhanov:1990me}.
}
\end{figure}

In the Fourier space, the transfer function for linear perturbations at wavenumber $k$ and time $t$ is defined as: 
\begin{equation}
T\left(k,t \right) \equiv \frac{\zeta_{\bf k}\left(t\right)}{\zeta_{\bf k}\left(t=0\right)}.
\end{equation}
This function, for typical cosmological parameters and evolved from the early universe 
to the present, can be seen in Figure \ref{fig:trans1}.

In order to determine what is passed from one cycle to the next, we need 
the transfer function at the infinite future:
\begin{equation}
T(k) \equiv \lim_{t\rightarrow \infty} T(k,t).
\end{equation}

The cosmic future appears to be 
dark energy dominated. If the structure of dark energy is known, we can evolve the 
same equations we used to calculate the transfer function to today, into the future. 
Furthermore, if dark energy is just a simple cosmological constant, the 
transfer function is not expected to evolve significantly after recombination, i.e. $T(k) \simeq T(k,t)|_{\rm present}$. This is because the speed of sound 
of matter is close to zero and dark energy is non-dynamical.

\begin{figure}
\includegraphics[width=0.40\paperwidth]{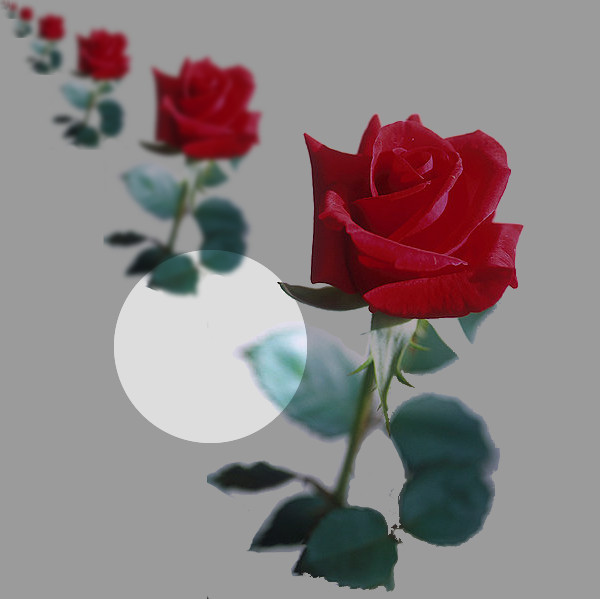}
\caption{\label{fig:flower}
An image depicting a figure repeated over several cycles, each time shrunk and then convolved with a low-pass filter (or transfer function), as in Figure (\ref{fig:trans1}). This is an example of a pattern that satisfies the matching consistency condition (\ref{matching}) in PTC. In cosmology, what we see is limited by the Hubble horizon, depicted here as the boundary between the light and dark regions. As such, only the light region of the universe would be visible to us.
}
\end{figure}

With this introduction, the consistency condition (\ref{matching_naive}) for primordial cosmological metric perturbations takes the form:
\begin{equation}
g_{ij}(\alpha {\bf x})|_{\rm big~bang} = T\star g_{ij}({\bf x})|_{\rm big~bang},\label{matching}
\end{equation}
where $ T\star$ denotes convolution with the linear transfer function $T(k)$ \footnote{See Sec. \ref{nonlinear}, for a nonlinear generalization of this matching condition.}. An example of this consistency under convolution+rescaling is depicted in Figure (\ref{fig:flower}). 

As convolution is a simple product in Fourier space, the primordial power spectrum (at the big bang) should satisfy the following simple consistency condition: 
\begin{equation}
{\cal P}(\alpha^{-1}k) = {\cal P}(k)|T(k)|^2.\label{power_matching}
\end{equation}
Given this condition, it is easy to see that the power spectrum must be of the form
\begin{equation}
\mathcal{P} \left( k \right)=F\left(\ln k\right) \prod_{n=1}^{\infty} \left|T\left(\alpha^n k\right)\right|^2, ~~{\rm for } ~~ \alpha < 1,
\end{equation}
where $F\left(\ln k\right)$ is periodic with a period of $\ln\alpha$: $ F\left(\ln k\right) = F\left(\ln k + \ln \alpha \right)$.
Here, for simplicity, we will just take $F\left(\ln k\right) = A_0$, a constant. This can be justified by requiring that on scales much bigger than all the physical length scales in the problem,  physics {\it must} be conformal and thus the spectrum must be scale invariant, not just periodic. We thus posit that {\it only} the matter transfer  function $T(k)$ breaks this symmetry.

Our resulting power spectrum is then  
\begin{equation}\label{eq:ps}
\mathcal{P} \left( k \right)=A_{0} \prod_{n=1}^{\infty} \left|T\left(\alpha^n k\right)\right|^2, ~~{\rm for } ~~ \alpha < 1.
\end{equation}
This function would be scale invariant if it were not for $T(k)$, as expected.
However, $T(k)$ looks very unlike what we expect for the primordial power spectrum. In addition, 
the infinite product seems to push the power spectrum to zero on all scales.
However, if we look at the large scales of $T(k)$, matter has very little effect, 
and it is nearly scale invariant. In addition, the tilt is red. Therefore, if we choose $\alpha\ll 1$, 
we will have a power spectrum close to what we expect. In addition, since 
$\alpha^{n}k\rightarrow 0$ for large $n$, given Equation \ref{eq:normT}, the 
infinite product should have a finite, non-zero value at the scales of interest 
since the terms get closer and closer to one.

\subsection{Approximating the Primordial Power Spectrum}

Now that we have a potential primordial power spectrum, we need to test if this can fit 
the data. To do this, we use CosmoMC \cite{Lewis:2013hha,Lewis:2002ah}. However, while CosmoMC is easy to modify 
for a new power spectrum, our power spectrum is not in an analytic form. Fortunately the 
program already calculates the transfer function to the present time. As we discussed above,
$T_{t\rightarrow\infty}$ is very close to $T_{present}$ for $\Lambda$CDM    cosmology. Therefore, 
we use $T_{present}\left(k\right)$ instead of $T_{t\rightarrow\infty}\left(k\right)$ 
for the transfer function in Equation \ref{eq:ps}.

Given $T\left(k\right)$, in order to calculate $\mathcal{P}$, we need to know the limit 
of the infinite product. We start with the approximation $\alpha \ll 1$. From 
Equation \ref{eq:normT}, we know that $T\left(\alpha^{n}k\right)$ in Equation \ref{eq:ps} 
will get closer and closer to one as $n\rightarrow \infty$. We will start with one factor of 
$\left|T\left(\alpha^{n} k\right)\right|^2$,
\begin{equation}\label{eq:ps1}
\mathcal{P} \left( k \right) \simeq A_{0} \left|T\left(\alpha k\right)\right|^2 , 
\end{equation}
and check whether our approximation is close enough.

We will start by determining if cutting off the product is a correct approximation. 
For the case when $\tilde{k} = \alpha^{n} k$ is sufficiently small, for 
$\tilde{k} \lesssim 10^{-4} \; \mathrm{Mpc}^{-1}$, from simple analyticity considerations, we expect the leading correction to the transfer function to look like
\begin{equation}\label{eq:Tlowk}
T\left(\tilde{k}\right) \simeq 1 - \epsilon \tilde{k}^2,
\end{equation}
where $\epsilon$ is the inverse square of an appropriate comoving sound horizon. In this case, Equation \ref{eq:ps} will become
\begin{equation}
\mathcal{P} \left( k \right)=A_{0} \left[ 1 + \epsilon^2 \alpha^4 k^4 - 2 \epsilon \alpha^2 k^2 - 2 \epsilon \alpha^4 k^2 + \mathcal{O}\left(\alpha^6\right) \right].
\end{equation}
This implies that to any finite order, Equation \ref{eq:ps} can be approximated by a 
finite product, and Equation \ref{eq:ps1} is correct to $\mathcal{O}\left(\alpha^3\right)$
for low $k$.

Now that we know a finite product of transfer functions will work, we estimate $\alpha$. 
We know our approximation, here Equation \ref{eq:ps1} is required to
maintain a small spectral tilt throughout the relevant scales. We know this is sufficient 
since further factors of $T$ will reduce the power of the smaller scales more than the 
larger ones, increasing the effective tilt. Using Figure \ref{fig:trans1}, our first 
estimate for $\alpha$ is for it to be on the order of $k_{\rm inflection}/k_{\rm max} \sim 0.01 $.
With this shift, the successive factors of $T(\alpha^n q)$ will quickly approach unity to 
the relevant degree of precision. For relevant $k$ ($k\le 1\; \mathrm{Mpc}^{-1}$), 
$\alpha^2 k < 10^{-4}$, which is 
within the range approximated by Equation \ref{eq:Tlowk}. In this case, each additional term corrects 
the previous by a term of order $\alpha^2 = 10^{-4}$ relative to the previous, which is 
less than $1\%$. This means we can approximate Equation \ref{eq:ps} by Equation \ref{eq:ps1}. 

However, the transfer function calculated by CosmoMC is calculated at only specific $k$ values 
calculated by the program. In order to use this with a shift in the values of $k$, we will 
need an interpolation between points and an extrapolation function for low $k$ values. 
For interpolation, we use linear interpolation. For 
extrapolation, we use Equation \ref{eq:Tlowk}.
We fit the smallest 30 $k$ values to the function using a least squares fit, 
and use it in place of any value of $k$ smaller than the 10th.

Now that we have our power spectrum, we need to estimate the value of $\alpha$ as well as 
the priors for this parameter. We will estimate this parameter by calculating the 
slope of $\mathcal{P}$ at the pivot scale $k_*=0.05\; \mathrm{Mpc}^{-1}$ and matching this to the 
$\Lambda$CDM    known $n_s=0.96$. Based on this our estimate is $\alpha \sim 0.007$.

We use as priors $0.0001 < \alpha < 0.1$. We do not expect that CosmoMC examines the 
whole range of this parameter space, however. Based on Figure \ref{fig:trans1}, we 
expect the error in Equation \ref{eq:ps1} to be $<1\%$ for $\alpha < 0.04$ for all the values 
of $k$ within our range.

\subsection{Beyond $\Lambda$CDM}\label{paramdescr}

Since our model uses part of the transfer function in place of the primordial 
power spectrum, we know that any effect which affects the shape of the transfer 
function will change the primordial power spectrum. Thus, the extensions of 
$\Lambda$CDM    which alter the shape of the transfer function will potentially be more 
relevant for our periodic model than they are for the standard power-law. Because of this, we 
will consider a variety of $7$ and $8$ parameter models in addition to the standard $6$ 
parameter model when evaluating the viability of PTC as presented here.

We consider two potential such effects, that of neutrinos and that of dark energy. 
We do not consider curvature as the conformal matching requires a flat universe. We also, as is typical, use the Big Bang Nucleosynthesis constraints to 
determine the helium fraction, so this will not be treated as a free parameter.
The parameters and their priors are in Table \ref{tab:prext}. 

\begin{table}
\caption{\label{tab:prext}
The priors used for the parameters extending the $\Lambda$CDM    astrophysics.
}
\noindent \begin{center}
\begin{tabular}{|c||c|c|}
\hline
 Parameter  & Minimum  &   Maximum   \tabularnewline
\hline
\hline
$N_\nu$         & $3$ & $4$ \tabularnewline
\hline
$m_{\nu}$          & $0.05$ & $0.5$ \tabularnewline
\hline
$w$             & $-1.1$ & $-0.9$ \tabularnewline
\hline
$w_a$           & $-0.1$ & $0.1$ \tabularnewline
\hline
$\Omega_\Lambda$           & $-0.01$ & $0.02$ \tabularnewline
\hline
\end{tabular}
\par\end{center}

\end{table}

Our choice of priors was chosen to prevent the numbers from getting too large. 
However, Planck has presented a range of potential values for these variables based 
on an analysis of the extended $\Lambda$CDM    models. We will leave for a future project a run CosmoMC 
with an extended parameter range based on the maximum range the Planck team analysis has suggested.

\subsubsection{Neutrinos}

There are two standard parameters in the extended $\Lambda$CDM related to neutrinos. These 
are the sum of the neutrino masses $m_\nu$ and the effective number of neutrinos $N_\nu$. 
Neutrino mass is typically taken to be minimum value possible based on neutrino oscillation experiments. 
This has a minor effect on the transfer function, but may effect the shape of the large scale spectrum 
which is relevant for our  PTC. 

The effective number of neutrinos is typically taken to be $3.046$, which is calculated from 
the standard model. This can also be taken to be a free parameter, calculated by fitting 
the data to the model. In this calculation, however, this parameter can be degenerate with $H_0$.

\subsubsection{Dark Energy}

Dark energy is an unknown factor in the evolution of the universe, and is commonly assumed to be a cosmological constant. 
However, since the effects of dark energy are going to be significant for our model, we 
will examine a few simple possibilities. 

Beyond a cosmological constant, the simplest popular alternative is when dark energy has a 
constant equation of state. In this case, the equation of state is defined by a constant $w$, 
where $p=w\rho$, with $p$ being the pressure and $\rho$ being the density. $w=-1$ corresponds 
to the case when dark energy is a cosmological constant, but variations of a few percent are still
consistent with current limits (e.g., \cite{planck2018cosmo}). 

Since we are using the current transfer function as our infinite future transfer function, 
the real form of $w$ we compare here is constant to now, then is set to $-1$ from now to 
the future. While this model is not what we expect, it will still determine if our model 
is viable when dark energy is not a cosmological constant. The case where we 
use the correct transfer function taken to the infinite future will have to wait for a future 
analysis (and possibly a more realistic model of dynamical dark energy). Moreover, a constant-$w$ model is a toy model to begin with, even if 
it were taken to the infinite future. 

To continue this examination, we also try a common extension to this model, 
$w_a$, which introduces a variation of $w$ linear in $a$, the size of the universe. If 
$w_a$ is taken alone without varying $w$, the present day $w$ would be set to $-1$.

As another alternative model, we use early dark energy (+ a cosmological constant). This could, for example, appear for exponential quintessence potentials \cite{Ferreira:1997hj}, or the quadratic cuscuton model \cite{Afshordi:2007yx}.
This dark energy component will affect the early times, but will 
become irrelevant later. That means that the transfer function still stops changing by 
the present, so it does not need to be evolved to the infinite future. 

The early dark energy component is assumed to be a constant fraction of the total density:
\begin{equation}
\Omega_{EDE} = \frac{\rho_{EDE}}{\rho_{tot}} = {\rm const.},
\end{equation}
with 
\begin{equation}
\rho_{tot} = \rho_{m}+\rho_{r}+\rho_{DE}
\end{equation}
being the total energy density. In this case, 
\begin{equation}
\rho_{DE} = \Lambda + \Omega_{EDE} \rho_{tot}.
\end{equation}
If we solve this for the equation of state, $w$, we get 
\begin{equation}
1+w_{DE} = - \frac{\dot{\rho}_{DE}}{3H\rho_{DE}} 
= K \frac{3\Omega_{0m}a^{-3} + 4\Omega_{0r}a^{-4}}{\Omega_{0DE}-\Omega_{EDE} + \Omega_{EDE}\left(\Omega_{0m}a^{-3} + \Omega_{0r}a^{-4}\right)},
\end{equation}
where
\begin{equation}
K = \frac{\Omega_{EDE} \left(1-\Omega_{EDE}\right)}{3}.
\end{equation}

%
%
%

\section{Comparing to Data}\label{data}

Now that we have our model set up, we can compare its predictions to data. As previously stated, we use 
CosmoMC in order to find a fit. In order to see how good of a fit this is, we will compare 
to the fit to the standard power-law primordial power spectrum given by 
Equation \ref{power_law}.

For both of these models and every choice of parameters to run, we find the best fit 
parameters as well as their means and expected ranges. We first run CosmoMC to 
get the distribution using a Monte Carlo Markov Chain, then run the minimizer included 
in the CosmoMC code to get the best fit parameters. For the case of the power-law power spectrum, we 
also compare the results we got with those of the Planck team to see if they make sense.

The datasets we use are those supplied by the Planck team. We used Planck 2015 as the Planck 2018 likelihood code has not yet been released, and when the analysis was run, Planck 2018 had not been released. For this, we include the 
$l \le 30$ temperature and polarization data (low TEB), the $l > 30$ temperature and polarization 
data (TTTEEE) as well as the lensing data \cite{Ade:2015xua,Planck2016i,Ade:2013nlj,Ade:2015zua,Aghanim:2015wva,Ade:2015fva,Ade:2013zuv,Planck:2013kta,Ade:2013mta,Bennett:2012zja,Das:2013zf,Reichardt:2011yv}. In addition, we use BAO \cite{Anderson:2013zyy,Beutler:2011hx,Ross:2014qpa,Anderson:2012sa,Padmanabhan:2012hf,Blake:2011en,Samushia:2013yga,Beutler:2012px}, 
the BICEP2-Keck-Planck joint dust analysis \cite{Ade:2015tva} and an instantaenous reionization at $z_{re} >6.5$ \cite{tauprior}. This last one 
was used since $\tau$ had been found to decrease for the best fit PTC model 
values.

To determine if our model is viable, we compare the best fit $\chi^2$ for the periodic time power spectrum
to the standard power-law power spectrum using the same cosmological parameters, with the sets of parameters described in Subsection \ref{paramdescr}. Our results are shown in 
Table \ref{tab:chi2}. As can be seen from 
Table \ref{tab:chi2}, PTC is ruled out at around $5\sigma$ for all $w=-1$ cases. This 
means our model requires non-trivial late-time dark energy.

\begin{table}
\caption{\label{tab:chi2}
Minimum $\chi^2$ values found for various runs of power-law  and periodic time cosmology (PTC) power spectra with the full dataset used here in the CosmoMC minimizer, as well as the difference between these $\chi^2$ values. If the value is found to be greater than that for one fewer parameters, the $\chi^2$ for the case of fewer parameters is listed. The sigma value is calculated by the square root of the difference. A model is expected 
to be ruled out above $3\sigma$. As can be seen, any combination of parameters which excludes varying $w$ puts the difference between models at $5 \sigma$, while including $w$ as a parameter puts the difference below $2\sigma$, causing the model to be viable.
}

\noindent \begin{center}
\begin{tabular}{|c||c|c||c|c|}
\hline
                  & $\chi^2$ for  power-law power spectrum    &  $\chi^2$  for PTC power spectrum  & $\Delta\chi^2$ &$\sigma$  \tabularnewline
\hline 
\hline 
$\Lambda$CDM       & $12995.3$ & $13022.0$ & $26.7$ & $5.2$\tabularnewline
\hline
\hline
+ $N_\nu$          & $12995.3$ & $13019.1$ & $23.8$ & $4.9$\tabularnewline
\hline
+ $m_\nu$          & $12995.3$ & $13019.0$ & $23.7$ & $4.9$\tabularnewline
\hline
+ $w \neq -1$      & $12995.3$ & $12998.6$ & $3.3$  & $1.8$\tabularnewline
\hline
+ $\Omega_{EDE}$   & $12993.9$ & $13020.9$ & $27.0$ & $5.2$\tabularnewline
\hline
\hline
+ $N_\nu$, $m_\nu$ & $12995.3$ & $13015.9$ & $20.6$ & $4.5$\tabularnewline
\hline 
+ $w$, $N_\nu$     & $12995.3$ & $12998.6$ & $3.3$  & $1.8$\tabularnewline
\hline 
+ $w$, $m_\nu$     & $12995.3$ & $12998.2$ & $2.9$  & $1.7$\tabularnewline
\hline 
+ $w$, $w_a$       & $12995.3$ & $12998.0$ & $2.7$  & $1.6$\tabularnewline
\hline 
\end{tabular}
\par\end{center}
\end{table}

Table \ref{tab:bf} shows the best fit and $68\%$ regions for both PTC and the standard power-law cosmology    
for the case when $w\neq-1$. From this we can see that the values for $w$ for PTC lie within the expected values from the standard power-law. However, PTC is much more sensitive to the value of $w$, causing it to have about a tenth of the range of values for $w$ within the listed $1\sigma$ range.

We can further see that the values of $\tau$ and $\Omega_c h^2$ decrease by about $1\sigma$, in the PTC model. 

\begin{figure}
\includegraphics[width=0.70\paperwidth]{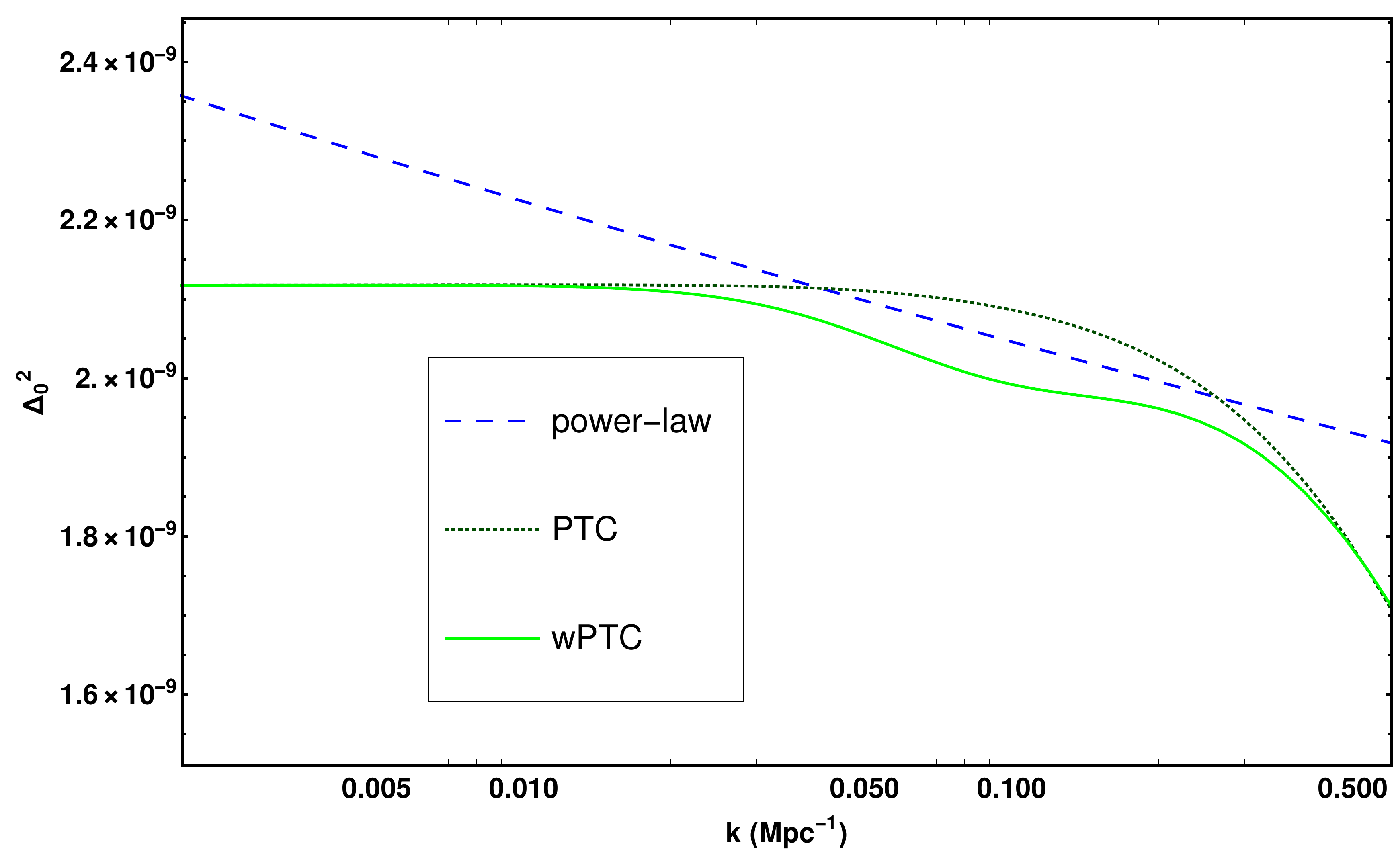}

\caption{\label{fig:pps}
Best fit primordial power spectra for power-law power spectrum (blue), periodic time cosmology (red, PTC+$\Lambda$CDM), and PTC with $w\neq-1$ and the transfer function evolved to the present (green, PTC+wCDM). The portion of the transfer function for PTC+wCDM near the pivot scale $k_* = 0.05$ Mpc$^{-1}$ can be seen to match the best-fit power-law. At larger scales, there is less power for PTC+wCDM compared to the power-law.
}
\end{figure}

A comparison of the primordial power spectra for a power-law, PTC with $w=-1$ and PTC with $ w \neq -1$ can be seen in Figure \ref{fig:pps}. Here we can 
see that $w \neq -1$ dark energy changes the shape of the primordial power spectrum, causing it to 
have a shape which resembles the tilt of the best-ft power-law model. In a future study, we will further examine this apparent correlation between $w$ and the effective tilt, and how it could change as we  evolve the transfer function into the future. 

\begin{figure}
\includegraphics[width=0.70\paperwidth]{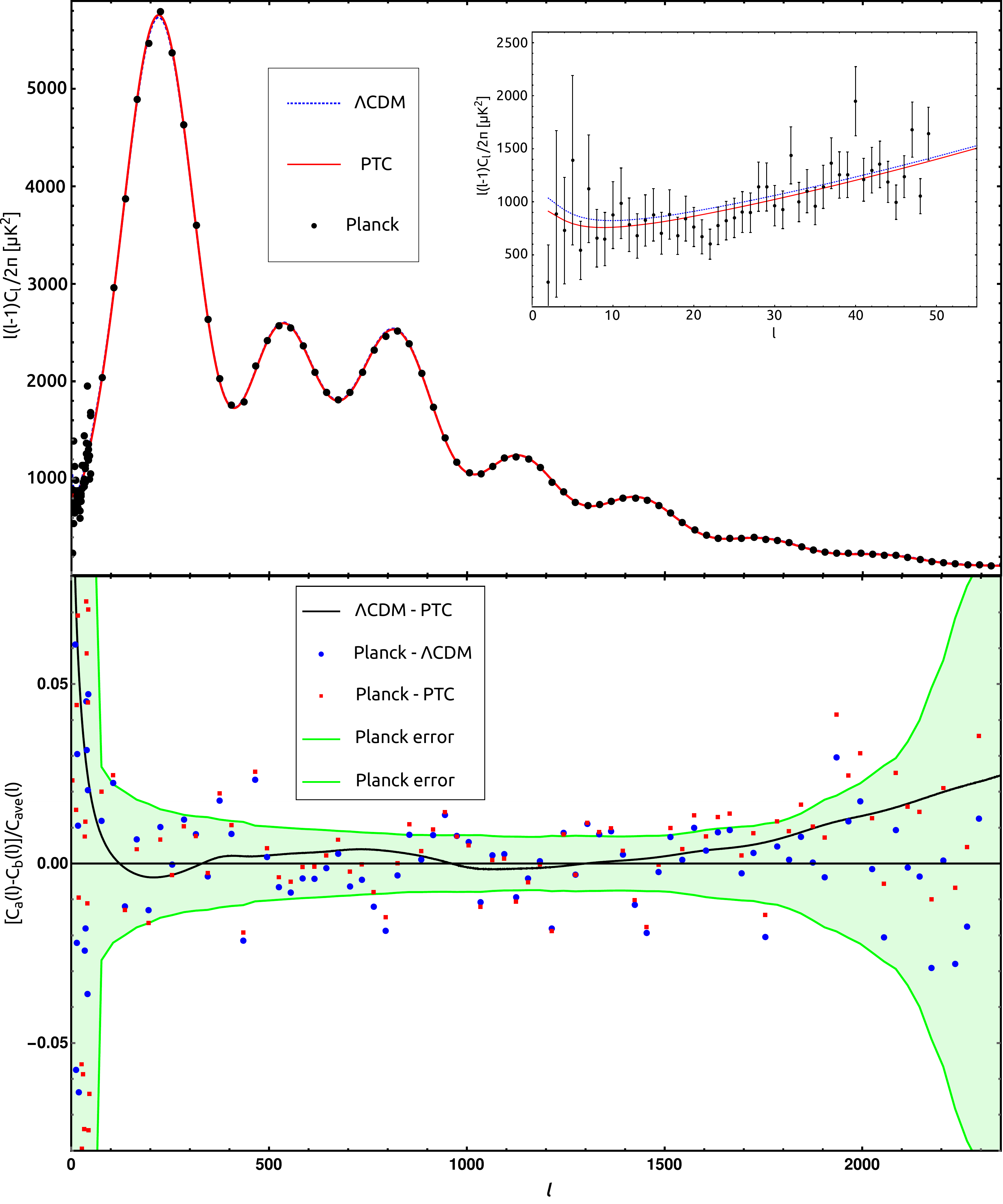}

\caption{\label{fig:tt}
A plot of the differences in temperature power spectra between power-law+$\Lambda$CDM    
and PTC+wCDM power spectra (see Figure \ref{fig:pps}) .
}
\end{figure}

\begin{figure}
\includegraphics[width=0.70\paperwidth]{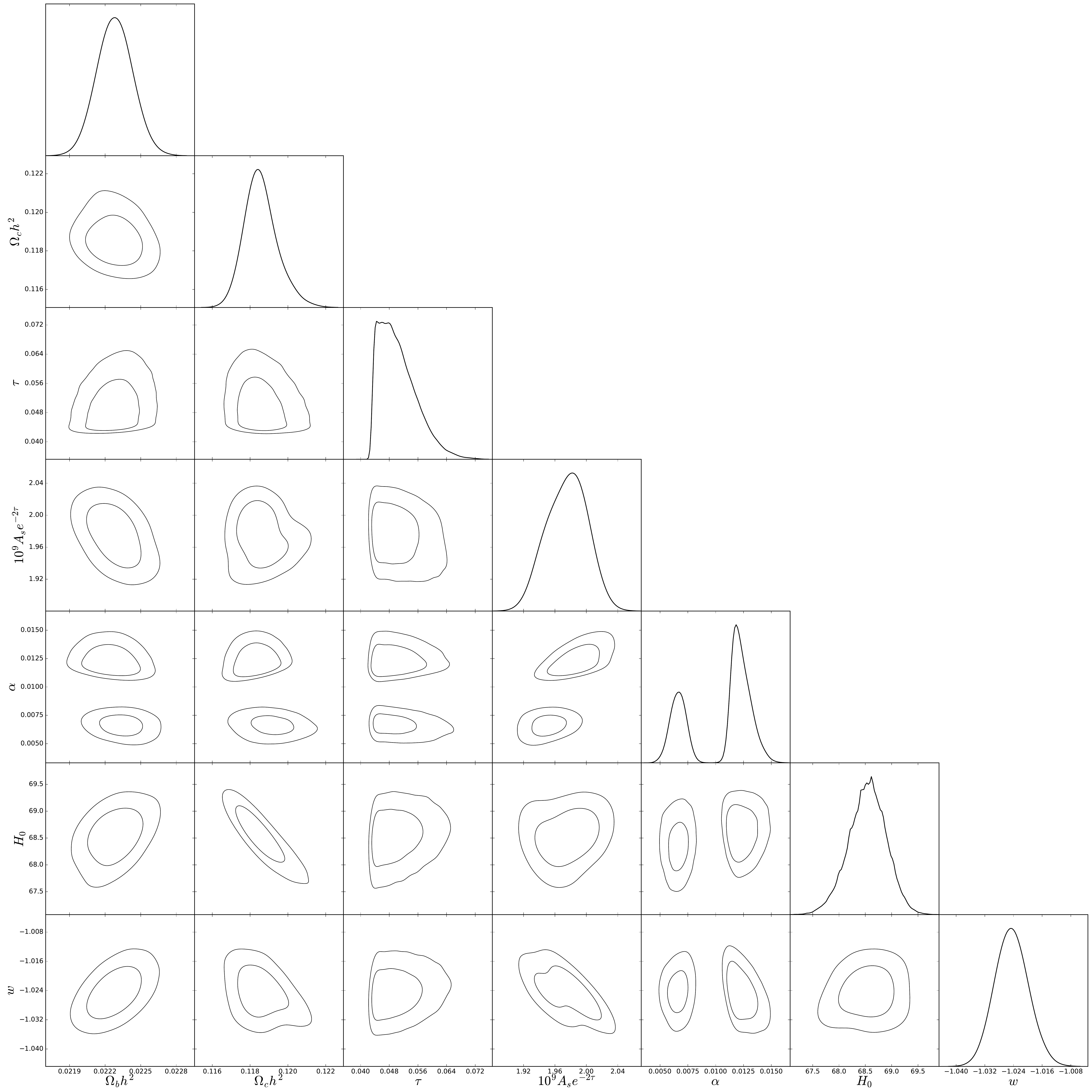}

\caption{\label{triplot}
A triangle plot of the parameter distributions for periodic time cosmology with non-trivial late time dark energy.
}
\end{figure}

For basic analysis of the results of the new model, the TT power spectrum for PTC with non-trivial dark energy can be seen 
in Figure \ref{fig:tt}. The fit seen here appears reasonably close to the Planck results. Figure \ref{triplot} contains the triangle plot of relevant altered parameters. We can see $\tau$ here reaching its prior as well as a double-peaked posterior for the new $\alpha$ parameter in the PTC model. 

\begin{table}
\caption{\label{tab:bf}
Best fit parameters and parameter distribution for periodic time cosmology (PTC) and power-law power spectra
for the case when $w\neq-1$.
}

\noindent \begin{center}

\begin{tabular}{|c||c|c||c|c|}
\hline
                  & \multicolumn{2}{c||}{power-law + wCDM} & \multicolumn{2}{c|}{PTC + wCDM}    \tabularnewline
\cline{2-5} 
                  & best fit  & $68\%$ range         & best fit  & $68\%$ range          \tabularnewline
\hline 
\hline 
$\Omega_{b}h^{2}$ & $0.02222$ & $0.02221\pm 0.00014$ & $0.02224$ & $0.02228\pm 0.00015$  \tabularnewline
\hline 
$\Omega_{c}h^{2}$ & $0.1198$  & $0.1197 \pm 0.0011$  & $0.1185$  & $0.1186 \pm 0.0009$   \tabularnewline
\hline 
$H_0$             & $68.2$    & $68.3   \pm 1.4$     & $68.54$   & $68.54  \pm 0.36$     \tabularnewline
\hline  
$\tau$            & $0.0566$  & $0.0546 \pm 0.0070$  & $0.0471$  & $0.0505 \pm 0.0053$   \tabularnewline
\hline 
$10^9 A_0$        & $2.110$   & $2.098  \pm 0.019$   & $2.184$   & $2.185  \pm 0.033$    \tabularnewline
\hline
$w$               & $-1.033$  & $-1.033 \pm 0.055$   & $-1.0248$ & $-1.0245\pm 0.0047$   \tabularnewline 
\hline
$n_{s}$           & $0.9639$  & $0.9639 \pm 0.0039$  &           &                       \tabularnewline
\hline 
$\alpha$          &           &                      & $0.0113$  & $0.0104 \pm 0.0028$   \tabularnewline
\hline 
\hline
$\chi^{2}$        & $12995.4$ &                      & $12998.6$ &                       \tabularnewline
\hline  
$mean(-Ln(like))$        &  &       $6514.9$      &  &       $6516.8$       \tabularnewline
\hline  
$-Ln(mean like)$        & &       $6509.7$      &  &       $6511.5$   \tabularnewline
\hline  
\end{tabular}

\par\end{center}
\end{table}

\section{Issues and Considerations}\label{discuss}

Now that we have seen that our model may fit the data reasonably well, 
we wish to ask what this fit means within the field of early universe cosmology. 
There are many aspects of periodic time cosmology to consider.

\subsection{Phenomenology vs. Theory}

As a phenomenological model, our model is based on testing the constraint of periodicity 
rather than developing an underlying theory. As such, it is a proof of principle that 
CTCs, often overlooked, could potentially provide a viable and consistent description of cosmic history. However, it does not give a reason for why the matching exists in the first place. 
It may be worth examining how this result fits in with CCC or the overall landscape of 
cyclic models, especially a microscopic mechanism for how such a matching could be produced in the context of a quantum theory of gravity.

\subsection{The Creation of a Power Spectrum}

Both inflation and the cyclic models will shift the small scale wavenumbers to larger 
scales, causing the visible power spectrum to be removed as the transfer function goes
to zero at small scales. For inflation, the universe is inflated at early times, while 
bounce models do this at late times (during dark energy domination). 
This leaves an empty background on which a new nearly scale 
invariant power spectrum is constructed by some mechanism dependent on the model. 

\begin{figure}
\includegraphics[width=0.70\paperwidth]{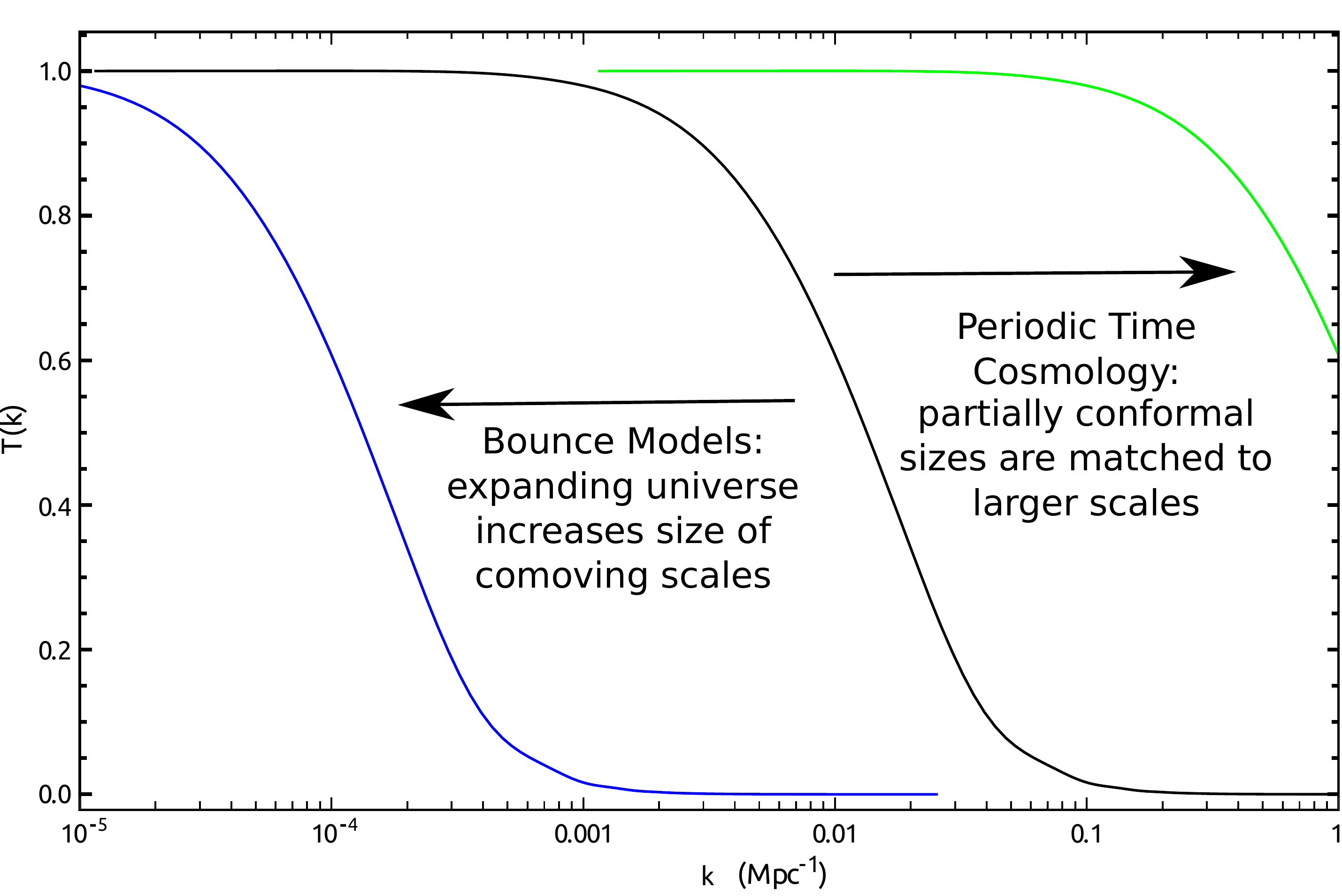}

\caption{\label{fig:transm}
A plot of the transfer function with the shift in comoving scales of matching for our conformal periodic time cosmology model vs. the standard inflationary or bounce model expansions depicted. Our periodic time cosmology model shrinks large scales into small scales and recycles its power through each cycle, while most other models wash away the existing power spectrum to reform a new one. The transfer function here is based off of Planck 2018 best fit values \cite{planck2018cosmo} as in Figure \ref{fig:trans1}.
}
\end{figure}

In part because of this trait, our model, despite using some of its conceptions, 
is not the same as Penrose's CCC. Not only was this 
original suggestion not exactly periodic, it also shifts the old power spectrum to zero and 
required a mechanism to reform the power spectrum as other cyclic models.
Our PTC model instead takes the shift of the power spectrum to be in the opposite direction.
The reason for this is to allow for the periodicity to form a constraint
on the new power spectrum instead of trying to conceive of a process
to construct a new one. 
The comparison of these two mechanisms is shown in Figure \ref{fig:transm}.

In our model, we reuse (or recycle) the primordial power spectrum and everything else in the universe. In
this way, it is unnecessary for us to create a mechanism for new matter to be created and a
new primordial power spectrum to form. 
Typically the generation of the primordial power spectrum is required to predict something, 
but consistency in the context of PTC is enough to make testable predictions, without any generation mechanism. 

\subsection{Cyclic vs Periodic Models}

Our model is designed on having exact periodicity to reduce the infinite series of universes
required for cyclic models. We then use this exactness to construct a method to constrain the 
form of primordial power spectrum based on the required exact repetition. However, as stated,
our model is phenomenological in nature, using the conditions of constraint to define the model
instead of using a specific model to calculate the results. It is possible to therefore discuss
the necessity of exact periodicity for this model to work. Without exact periodicity, this
model would no longer have the conceptual niceties of not having a series of universes or 
the perfect constraint from our proposed model. 
However, the constraints would be loosened and the issues with having closed timelike curves
would not be relevant.

A potential future examination of this may look at if this model is an attractor to determine
if a series of universes would eventually lead to this type of thing.

\subsection{Dynamical Dark Energy?}

During our analysis, we found that non-trivial dark energy was required for our model 
to work. However, this may not be surprising. Modern cyclic universe models rely on the 
late time behavior of the universe. And CTCs typically require matter violating the energy 
conditions to exist. Future precision tests of dark energy history (e.g., WFIRST, Euclid, LSST) will be able to test this prediction. While the best-fit PTC+wCDM model studied here, does not significantly help with current tension with local measurements of Hubble constant \cite{Riess:2016jrr, Birrer:2018vtm}, it is clear that a more general dynamical dark energy model has the potential to do so (e.g., \cite{Zwane:2017xbg}).

\subsection{Beyond Linear Perturbation Theory}\label{nonlinear}

While we have focused on cosmological linear perturbations throughout the paper, the setup can be easily extended beyond linear theory.  To see this, note that in the matching condition (\ref{matching}), we can simply replace the linear transfer function by the full nonlinear transfer function. This can be order-by-order in perturbation theory. Given that adiabatic cosmological initial conditions can be exclusively written in terms of scalar and tensor perturbations, $\zeta$ and $h_{ij}$, the 2nd order transfer function for $\zeta$ is given by:
\begin{eqnarray}
&&\zeta_{\bf k} \to T^s_{s}(k) \zeta_{\bf k} + \nonumber \\
&&  \int \frac{d^3{\bf k'}}{(2\pi)^3} \left[T^{ss}_{s}(k,k',{\bf k \cdot k'}) \zeta_{\bf k'} \zeta_{\bf k-k'} + T^{st}_{s}(k,k',{\bf k \cdot k'}) k'^i k'^j\zeta_{{\bf k'}} h_{ij,{\bf k-k'}}+ 
T^{tt}_{s}(k,k',{\bf k \cdot k'}) h^{ij}_{\bf k'} h_{ij,{\bf k-k'}} \right] + ... ~ ,\nonumber\\
\end{eqnarray} 
where $T^s_s$ is the scalar linear transfer function used in Sec. (\ref{ss:ppsd}) above, while $T^{ss}_s, T^{st}_s$ and $T^{tt}_s$ would come from 2nd order perturbation theory (e.g., \cite{Bartolo:2005kv}). A similar expression (but with richer tensor structure) can be written for the tensor modes.  

Now, plugging these expressions into the matching condition (\ref{matching}) would generalize the consistency condition for the power spectrum (\ref{power_matching}) to include bispectra, as well as 2nd order terms in power spectra of tensor and scalar modes. Therefore, in principle, the effects of nonlinear evolution in previous aeons can be included in the PTC power spectrum (\ref{eq:ps}) self-consistently, order by order. However, in practice, due to the rescaling at big bang, these effects should only show up on comoving wavenumbers with $ k \gtrsim \alpha^{-1}k_{\rm NL} \sim 10^2 k_{\rm NL} $, where $k_{\rm NL} \sim {\rm Mpc}^{-1}$ is the comoving scale significantly affected by nonlinear evolution in our current aeon.  In other words, nonlinearities in previous aeons will be irrelevant for predictions on all (quasi-)linear scales that are currently used for precision cosmology. 

A related issue would be other potential non-perturbative effects that could happen over exponentially long times, such as vacuum instability through quantum tunnelling \cite{1980PhRvD..21.3305C}. However, given that the comoving Hubble volume is shrinking as  $\exp(-3H t)$ in the near-de Sitter phase, these effects will be limited to exponentially small comoving volumes, and thus will have negligible statistical effect on cosmological observables. More generally, as in the case of inflationary reheating, nonlinear late-time processes on sub-horizon scales are expected to have little effect on super-horizon adiabatic modes that are already frozen (e.g., \cite{Liddle:1999hq}).

Therefore, we conclude that our predictions for PTC model are robust to nonlinear and/or non-perturbative corrections beyond linear perturbation theory (in previous aeons), at least given the current and foreseeable precision of cosmological observations.   

{
\subsection{Matter across the Transition}

While we have focused on the matching of geometries from infinite future to big bang, one may wonder about the fate of matter across the transition. 
During the late-time de Sitter phase, all the matter fields (including photons and neutrinos) are redshifted exponentially and thus have negligible impact on geometry as $t \rightarrow \infty$. Similarly, during the hot big bang as $t \rightarrow 0$, all the matter degrees of freedom are determined by local temperature, if one assumes adiabatic initial conditions. Therefore, only geometric degrees of freedom survive on either side of the transition, which is why they can be matched using  Equation \ref{matching}, without worrying about the matter content of the universe. Turning this around, this also implies that only {\it adiabatic} initial conditions are predicted in PTC cosmology.  

More concretely,  as Penrose indicated \cite{ccc}, matter will eventually mostly all fall into black holes. These black holes will then evaporate \cite{Hawking}, leaving only radiation at late times. However, given that this evaporation only happens at very late times (e.g., $10^{60}$ times the current age of the universe for stellar black holes) it can only impact extremely small scales in PTC cosmology. Moreover, the energy in these gravitational waves will still decay to zero in the late-time de Sitter phase. 

}

\subsection{Closed Time Considerations}

One of the big issues with this type of model is the consideration of the second law of
thermodynamics. This may or may not be a real issue, as some have noted the fact that 
this may only apply in an closed system and not when considering the full universe \cite{jb}. Throughout
previous sections, we have assumed this is not an issue. But let us consider here other solutions
to this problem for cyclic universes. The original cyclic models had issues due to increasing of
entropy at each cycle (which extended each cycle) \cite{Tolman}. The ekpyrotic model solves this issue by 
increasing the size of the universe each time, keeping a constant entropy density despite an 
increasing entropy \cite{ekpyro}. CCC, the model closest to ours, uses black hole information loss to 
remove the increasing entropy \cite{ccc}. Since our model decreases sizes instead of increasing them, 
no trick of increasing sizes to keep entropy density constant will work. The black hole case 
can work still, however. 

Having an infinite space may prevent the entropy considerations from being relevant \cite{jb}. This 
would be required in order to have conformal symmetry at any time, as a finite spatial volume would 
create a preferred scale, removing the conformal symmetry required to make this work. 

In talking about entropy, we left out consideration for the formation of an arrow of time.
Having a time dimension which is not identical to the spacial ones is already known, both by 
having this dimension be the finite one as well as by the usage of FRW spacetime. FRW has a
preferred time direction given that it lacks symmetry in that direction while maintaining symmetry
for spacial directions. Future can be given by the direction of increasing universe size, 
which has relevance during times without conformal symmetry. While this doesn't solve all 
problems, anything further is beyond the scope of this paper and probably unnecessary for our 
analysis.

\subsection{Other Properties to Find}

We also wonder if there are other observable signs of our proposed model. One 
proposal would be that there would be exact copies, and therefore an exactly 
fractal structure in the universe. Figure \ref{fig:flower} shows what an 
image would look like if it were rescaled each cycle and passed through a 
low-pass filter like matter transfer function. If the universe were periodic, the fractal 
pattern produced would need to be exact. The rescaling would also create a 
preferred point. It might be possible (but not guaranteed)  to find signatures of these repeating re-scaled structures in the large scale structure of the universe.

\section{Conclusion}\label{conclude}

In this paper, we developed a cosmological framework for periodic time dimension. This relies on the idea of matching infinite future and big bang, via conformal re-scaling, following the original suggestion by Penrose  \cite{ccc}.  
We show that consistency fixes primordial power spectrum in terms of cosmic history, in the context of periodic time cosmology (PTC).  

We then compared PTC predictions for standard cosmological histories to our current 
cosmological observations. We found that such a model can fit observations reasonably well. While the data still favors the standard power-law power spectrum (at 1.8$\sigma$ level), PTC model remains viable if  a non-trivial dark energy equation of state is considered. 

In our analysis, we found that having dark energy which is just a cosmological constant does 
not  reproduce observations, being ruled out at $5.1\sigma$. 
A further analysis indicates that the late time 
behavior for dark energy is important. 
These models are designed to take what will happen in the infinite future 
of our universe as a replacement for the effects which traditionally rely on inflation in 
the early universe. 

Many questions still remain. We need to evolve the transfer 
function into the infinite future to fit the data appropriately, possibly with a better-motivated dynamical dark energy. Beyond this, we hope to 
 find a more physical origin of the matching condition. 
Another issue that needs to be addressed is the status of the 2nd law of thermodynamics.

However, we end with one final question: Is it possible that all we need to recreate our past is to recycle our future?

\section*{Acknowledgements}
This work has been partially supported by the National Science and Engineering Research Council (NSERC), University of Waterloo, and Perimeter Institute for Theoretical Physics (PI). Research at PI is supported by the 
Government of Canada through the Department of Innovation, Science and Economic Development Canada and by the Province of Ontario through the Ministry of Research, Innovation and Science.

We acknowledge the use of the Legacy Archive for Microwave Background Data Analysis (LAMBDA), part of the High Energy Astrophysics Science Archive Center (HEASARC). HEASARC/LAMBDA is a service of the Astrophysics Science Division at the NASA Goddard Space Flight Center.

We acknowledge the use of the IRIDIS High Performance Computing Facility, and associated support services at the University of Southampton, in the completion of this work.

Computations were performed, in part, on resources and with support provided by the Centre for Advanced Computing (CAC) at Queen's University in Kingston, Ontario. The CAC is funded by: the Canada Foundation for Innovation, the Government of Ontario, and Queen's University.

\bibliographystyle{JHEP.bst}
\bibliography{PT}

\end{document}